\documentclass[fleqn]{article}
\usepackage{espcrc2}
\usepackage{graphicx}
\usepackage{amssymb}

\begin{document}

\title{An apparatus to control and monitor the para-D$_2$ concentration in 
a solid deuterium, superthermal source of ultra-cold neutrons}

\author{C.-Y. Liu\address[Princeton]{Physics Department, Princeton University,\\
	Princeton, NJ 08544}%
        \thanks{Corresponding author. P-23, MS H803, Los Alamos National Laboratory, Los Alamos, NM 08545, USA. Tel:505-665-9804; fax:505-665-4121; e-mail: cyliu@lanl.gov},
        S.K. Lamoreaux\address[LANL]{Physics Division,
        Los Alamos National Laboratory, \\
        Los Alamos, NM 08545, USA}, A. Saunders\addressmark, 
	D. Smith\addressmark[Princeton]%
	\thanks{current address: Stanford Linear Accelerator Center, Stanford, CA 94305, USA}, 
	and
        A.R. Young\addressmark[Princeton]%
	\thanks{current address: Physics Department, North Carolina State
	University, Raleigh, NC 27695, USA}
	}

\date{\today}

\maketitle

\begin{abstract}
Controlling and measuring the concentration of para-D$_2$ is an essential step
toward realizing solid deuterium as an intense ultra-cold neutron (UCN) source.  To this end, we
implemented an experimental technique to convert para- to ortho-deuterium molecules
by flowing D$_2$ gas through a cryogenic cell filled with paramagnetic hydrous ferric oxide
granules. This process efficiently reduced the para-D$_2$ concentration from 33.3\% to 1.5\%.
Rotational Raman spectroscopy was applied to measure the residual para-D$_2$ contamination
to better than 2 parts in 10$^3$, and the hydrogen contamination to 1 part in 10$^3$.
We also contrast our optical technique to conventional thermal conductivity
measurements of the para-D$_2$ concentration, reporting some of the relevant
strengths and weaknesses of our implementation of each technique.
\end{abstract}

\section{Introduction}

In a recent investigation of solid D$_2$ as a UCN source\cite{Serebrov96}, it was  
surmised that the yield of UCN is correlated with the spin states of 
the deuterium molecules. Subsequently, detailed neutron scattering calculations and
measurements of the UCN lifetime in mixtures of para- and ortho-deuterium
quantitatively verified this conjecture.\cite{Liu00,Los_Alamos,Morris01,Young00,Hill} 
The work we report here was motivated by our attempt to control the spin state 
of deuterium 
and develop a superthermal ultracold neutron source capable of the 
extremely high UCN
yields predicted for pure ortho-D$_2$ crystals by Golub $et\ al.$\cite{Golub83,Yu86}  At least three solid deuterium sources are currently under
construction.\cite{Morris01,munich,psi}

Our superthermal solid D$_2$ source at Los Alamos Neutron Science Center (LANSCE)
can be envisioned as roughly one liter of
solid deuterium held at temperatures below 10 K.  This block of deuterium
is then immersed in a cold neutron flux.  UCN are produced when the cold neutrons
are downscattered, losing almost all of their energy.  Because the source is
operated at very low temperatures, the loss rates due to thermal upscattering
are very small for UCN and the produced UCN can be efficiently extracted from the source.
The ultimate UCN densities one can achieve with a superthermal source 
are\cite{Golub83}:
\begin{equation}
\rho=P\tau,
\label{ucndensity}
\end{equation}
where  $\rho$ is the UCN density, $P$ is the downscattering rate for cold neutrons,
and $\tau$ is the UCN survival time in the solid D$_2$.  With a careful design,
the downscattered neutrons are prevented from 
subsequent scatterings within the source to establish thermal equilibrium; several
orders of magnitude improvement in available UCN densities are expected from these
sources. It is also clear from Eq. \ref{ucndensity} that longer UCN
survival times, $\tau$, equate to higher UCN densities.  Para-deuterium, if
present in the source, drastically increases the UCN loss rate and, therefore, decreases
$\tau$ and reduces the yielded UCN densities.  An essential key to optimizing the
performance of a solid D$_2$ UCN source is thus to minimize the concentration of
para-D$_2$ in the deuterium.  In what follows, we present
the results obtained during the first nine months of operation of our UCN source coupled to a para-to-ortho
converter.  Because we present the results of a development project, the performance of this
system was not fully optimized.  Nevertheless, the experimental technique is robust; it 
should form the basis of a workable approach to the operation of a solid deuterium
superthermal source of UCN.

In addition, HD, a common contaminant in commercially available deuterium, also greatly
reduces the survival time for UCN in our solid deuterium source, simply because of
the large neutron absorption cross-section of the hydrogen nucleus.  Ideally, our
experimental technique for monitoring the para-deuterium concentration
in our source should also be useful in monitoring HD levels. As we shall 
demonstrate,
this is the case for the Raman spectroscopy technique presented in this work.

The physics of hydrogen and deuterium solids has been thoroughly discussed
in various reviews\cite{Silvera80,VanKranendonk82,Farkas35}, so we
will merely touch on the relevant properties of these crystals.  
First of all, the deuterium molecule (D$_2$) has
a symmetric wave-function under permutations of the two identical nuclei, deuterons,
having a nuclear spin $I=1$. As a result,
the ortho state for deuterium ($I_{total}=0, 2$) can only have
symmetrical molecular rotational states of even $J$, the rotational quantum number, i.e., $J=0,2,...$; and
the para state ($I_{total}=1$) can only have anti-symmetrical
molecular rotational states of odd $J$, i.e., $J=1,3,...$.
At the typical low operational temperatures of superthermal sources, 
because of the relatively large vibrational excitation energy (first
vibrational excitation energy level is $\sim$4000K), all
of the molecules are expected to be in the vibrational ground state. On 
the other hand, it is possible to find significant populations of 
the first ($J=1$) rotational excited states of the molecules (para-D$_2$), 
in addition to the
ground state with $J=0$ (ortho-D$_2$), depending on how the sample
is prepared.    

\begin{figure}
\begin{center}
\includegraphics[width=2.2 in]{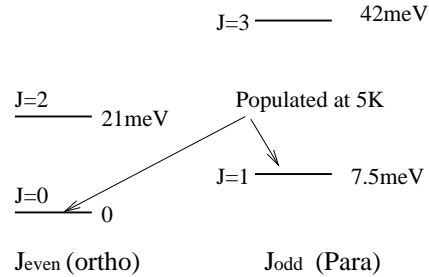}
\end{center}
\caption{The rotational energy spectrum of ortho- and para-D$_2$.}
\label{figu_Estate}
\end{figure}

The rotational energy spectrum (shown in Fig. \ref{figu_Estate}) of the D$_2$
molecule can be described using a classical dumbbell picture:
\begin{equation}
E_{rot}= \frac{J^2}{\mathbb{I}}=\frac{B}{2} J(J+1),
\end{equation}
\noindent where the factor $B$ is related to the momentum of inertia, $\mathbb{I}$.  
To a first approximation, 
$B=\hbar^2/\mathbb{I}=\frac{\hbar^2}{2M_{D}(a/2)^2}=7.5$ meV, in which   
the separation between the two nuclei, $a$, is 0.74 $\mathrm{\AA}$.  
The energy required to make the transition from the ground state ($J=0$)
to the first excited state ($J=1$) is thus 7.5 meV.

In the high temperature limit all the nuclear spin sub-states are equally populated.
This leads to an equilibrium para state concentration of 33.3\%.
For a free non-interacting gas, transitions between even $J$ (ortho) 
and odd $J$ (para) states are prohibited by the spin selection rules.
Conversion of para to ortho states involves a nuclear spin flip, which requires,
in general, an external interaction potential. In real situations, 
spontaneous spin flipping
occurs through 
collisions between the aspherical molecules ($I_{total}=1,2$), whose non-central
quadrupole interaction provides a small, yet finite torque to induce spin flip.
The resulting time constant for such a self-conversion is, however, very 
long, and has been measured to be on the order of months\cite{Silvera80}.

\section{Conversion of Para-D$_2$ into Ortho-D$_2$}

Our experimental goal is to reduce the para-D$_2$ fraction present in our cryogenic
solid D$_2$ source material to the level where it no longer limits the UCN
survival time, $\tau$, in the crystal. This can be achieved by reducing the para-D$_2$
concentration below about 1\%.  At this level, the UCN loss rate in the crystal
due to UCN upscattering from the para-D$_2$ is roughly the same as the
loss rate due to nuclear absorptions by the deuterium nuclei.\cite{Liu00}  Because
the neutron absorption losses in the deuterium itself are unavoidable,
little improvement can be obtained by further lowering the para-deuterium concentration.

An effective way to reduce para state concentrations is to pass D$_2$ gas through
a system which permits the D$_2$ to rapidly come into thermal equilibrium
at a low temperature by inducing a spin flip process.  The residual population of 
the $J=1$ state can be described by the Boltzmann factor, which is 
characterized by the transition energy, $E_{JJ'}=E_{01}=7.5$
meV($\sim$ 80 K). 
When the converter is operated at 17K, the para concentration could in principle
be reduced from 33.3\% down to $e^{-80/17}$ = 1\%. Converter temperatures below 17K 
can in principle give a smaller population of para states, but they
supply converted D$_2$ with less than the required vapor pressure 
(30 to 50 Torr) from the converter to sustain 
ortho-deuterium solid growth in a timely, continuous fashion in our application.

\subsection{Catalyzed Conversion}

We apply paramagnetic conversion\cite{Eley48,Selwood51} to prepare  
our deuterium samples.  The paramagnetic catalyst agent has
a large magnetic field near its surface, on 
which D$_2$ molecules are temporarily adsorbed and form a Van der Waals layer. 
The magnetic field gradient induces a relative
dephasing between the two precessing nuclear spins, and results in a spin flip.
For a single adsorbed molecule this transition rate is nearly temperature independent, 
and thus the conversion rate is dominated by  
the gas adsorption probability. Combining these two effects, one can predict that
the overall conversion efficiency increases as the catalyst temperature decreases, compatible with our low temperature application.

One of the most effective paramagnetic catalyst materials is hydrous ferric oxide, Fe$_2$O$_3 \cdot x$(H$_2$O), a self-supporting metal oxide in the form of colloidal gels.\cite{Weiser35} It has an effective surface
area of 100$\sim$300 m$^2$/g.\cite{Buyanov60,Weitzel55,Weitzel58}
As suggested by  
Ref.\cite{Buyanov60}, we carried out the initial activation 
by heating the catalyst up to 130 $^{\circ}$C for 24 hours inside the converter cell
under the vacuum and then back-filled the cell
with deuterium gas for preservation.  Installation of the converter cell into the
vacuum system was performed with care to minimize exposure to the ambient air.
After some period of operation, which varied with the conversion load, regeneration of
the catalyst was performed with the same activation procedure.

\subsection{Design of the Converter}

\begin{figure}
\begin{center}
\includegraphics[width=2.7 in]{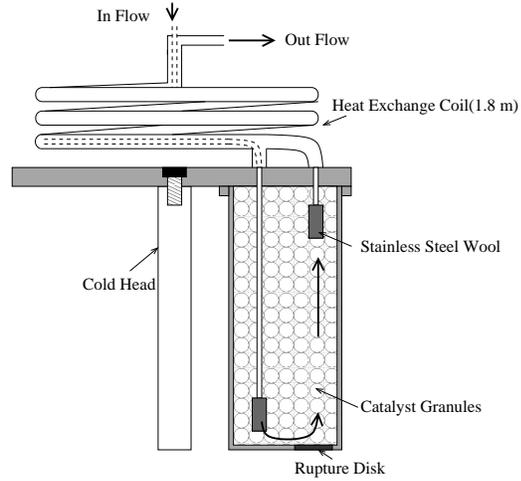}
\end{center}
\caption{A schematic of the para/ortho-D$_2$ converter.}
\label{figu_converter}
\end{figure}

We built the converter chamber out of a cylindrical cell made of OFHC copper,
illustrated in Fig.~\ref{figu_converter}. 
It is attached to the cold head of an
air-cooled $^4$He compression cryopump, Varian CryoTorr 8, which served as a refrigerator.
The cryopump has an inner cooling stage (cold head) with a cooling power of 
1 W at a temperature of 10K
(5W at 20K), which is adequate for our applied conversion load.
This closed system of $^4$He gas compression cryopump provides an easy and compact 
alternative to the liquid $^4$He evaporation cooling used in previous experiments\cite{Bazhenov}.

The converter cell's design features are summarized in the following: 
the D$_2$ gas exit port is situated at the coolest plate
to anchor the temperature of the output converted D$_2$; it also minimizes the
temperature gradient across the D$_2$ gas flow.   
In addition, in order to reduce the heat load
onto the cold head, we designed a 180 cm long heat exchange coil made of 
coaxial stainless steel (S.S.) thin tubes.  
The inflow gas inside the inner thin tube is precooled by the 
outflow gas flowing 
in the outer tube.  Its long length increases the thermal resistance and 
thus minimizes the
heat conduction through the S.S. tubes to the ambient environment.
With a flow rate of 36 liter/hour (0.6 liter/min) of STP gas, the heat load is 
about 1W. The heat load is dominated by contributions from  
the D$_2$ gas specific heat and the ortho/para-D$_2$ conversion energy. 
The 100 c.c. inner volume of the cell is filled with fine grained catalyst, 
which is confined in the converter cell by S.S. wool filters
plugged into both the gas entrance and the exit ports. 
A heater is mounted on the cold plate to provide us the capability to 
control and vary the temperature of the converter. Two temperature diode
sensors are separately placed on the top and the bottom of the copper cylinder to monitor the
temperature, as well as the gradient.  Moreover, in order to tolerate  
baking for catalyst regeneration, an In/Pb alloy wire is used between the top plate and
the converter can for the vacuum seal.  
This soft alloy has a higher melting point 
(150$^{\circ}$C) than the regeneration temperature (130$^{\circ}$C). Pure 
Indium metal (with a melting point of 110$^{\circ}$C) is not compatible with
regeneration baking.
We also established that these regeneration bake-outs could occur with the cell still
mounted to the cold-head of our cryopump, making regeneration of the catalyst
quite straightforward.

The operation of this converter is in a continuous mode described in the following. The pressure regulated 
D$_2$ gas is introduced into the converter; after 
flowing through the converter in which the catalyst is held below the D$_2$ triple
point temperature (18.7K), D$_{2}$ is converted predominantly into the ortho state with a
reduced para contamination of roughly 1\%. The converted D$_2$ is then 
continuously cryopumped onto the superthermal source cryostat (cooled by
liquid He).  We observed that it was possible to grow a
very transparent, large solid block of D$_2$ directly from flowing gas into our source
cryostat when the vapor pressure was kept below about
50 torr. The corresponding low D$_2$ flow rate is limited to acceptable levels 
at which the heat flow into the low temperature end of the cryostat
and the growing crystal.

\begin{figure}
\begin{center}
\includegraphics[width=2.9 in]{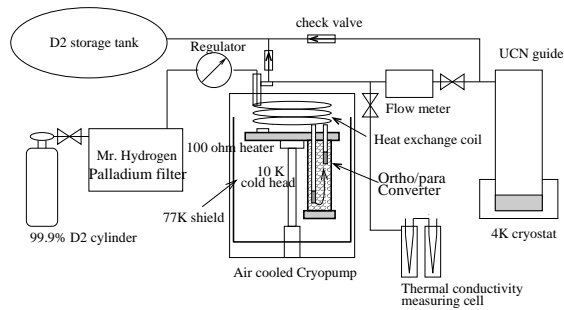}
\end{center}
\caption{A schematic of the gas handling system used for the S-D$_2$ 
UCN source preparation.}
\label{figu_gashandling}
\end{figure}

As illustrated in Fig. \ref{figu_gashandling}, this para-to-ortho-D$_2$ converter with
the $^4$He compression cryostat is incorporated into the D$_2$ gas handling system, which is used to prepare the UCN source. 
Cajon VCR S.S. tubes and connectors were used
to the maximum extent possible to construct a clean gas handling system.  The
para-to-ortho converter was put in the downstream of a hydrogen gas 
purifier (palladium filter)
which operates at 400 $^{\circ}$C.  Care was taken to minimize the magnetic surface area
inside the apparatus placed after the para-to-ortho-D$_2$ converter, in order to prevent
back-conversion and to preserve the fully converted ortho-D$_2$. Our measurements show 
that when converted D$_2$ gas is passed through our room temperature S.S. tubes and
Ni-coated UCN guides and then stored as a solid in our cryostat at low temperatures, we
did not observe significant back-conversion over the course of the experiment (typically two days).
Details of our UCN source itself can be found elsewhere\cite{Hill}.

\section{Measurement of the Ortho/Para-D$_2$ Ratio Inside the UCN Source}

We used two different, complimentary measurements to determine the para-D$_2$ 
concentration
in our UCN source.  An in-line thermal conductivity measurement with hot wire cells 
was frequently performed on the outflowing D$_2$ to characterize the 
performance of the para-to-ortho-D$_2$ converter.  Molecular Raman spectroscopy
\cite{Silvera80} was later conducted on the evaporated D$_2$, which was collected
from the cryostat at the end of each operational run of our source.

\subsection{Raman Spectroscopy}

This optical method uses a visible (far off-resonance) laser to probe 
D$_2$ molecule gas sample. The 
rotational Raman spectrum gives information on rotational state populations.
For example, for molecules in thermal equilibrium, the total Raman cross-section is
formulated as\cite{Penney74}
\begin{eqnarray}
\frac{d\sigma^{raman}}{d\Omega'} &=& \sum_J Q^{-1}g_{J}(2J+1)e^{-E_J/kT} \times \nonumber \\
&& \frac{d\sigma_{J 
\rightarrow J \pm 2}}{d\Omega'},
\label{eq_ramansum}
\end{eqnarray}
\noindent with `$+$' standing for Stokes lines, and `$-$' for anti-Stokes lines,
$Q$ is the partition function for the rotational states, and $g_J$ is the multiplicity of the  
nuclear spin states.  In the equation above we took $\frac{d\sigma_{2 
\rightarrow 4}}{d\Omega'}=1.3\times 10^{-30}$cm$^{2}$ from the absolute
rotational Raman cross-section for D$_2$ using the cross-sections quoted for
the $J=1 \rightarrow 3$ transition in H$_2$ at room temperature in Fenner\cite{Fenner73}.  We adjusted the cross-sections in Fenner for the difference in the initial spin states
and the  rotational state population differences between H$_2$ and D$_2$ gas,
and otherwise assumed the same anisotropy in the molecular polarizability
tensor.  This cross-section yields a consistent, order of
magnitude estimate of our measured count rates when the appropriate gas densities,
efficiencies and imaging characteristics for our system are applied. 

\begin{figure}
\begin{center}
\includegraphics[width=2.8 in]{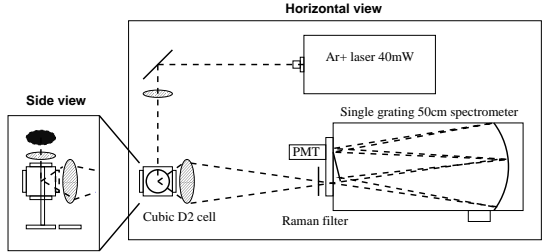}
\end{center}
\caption{A schematic of the D$_2$ Raman spectroscopy setup.}
\label{figu_Ramansetup}
\end{figure}

Our experimental implementation of rotational Raman spectroscopy is similar to the
approach discussed by Compaan and Wagoner\cite{Compaan94}. The setup schematic is 
shown in Fig. \ref{figu_Ramansetup}. 
We use an air-cooled CW argon ion laser, American Laser Corp. model 60x, which outputs
a single line of unpolarized 488 nm TEM$_{00}$ beam with a power of 40 mW, as the 
light source. The laser line is directed and focused by
two convex lenses ($fl$ = 35 mm) into the center of a cubic sample cell, 
which is built out of
a S.S. housing with 4 flat parallel uncoated quartz windows. A 58 mm Nikon camera
lens ($fl : \infty \sim 0.5$) is placed close to the cell to collect roughly 5\% of the
total solid angle of the scattered light. It also images the Raman 
radiation onto the entrance slit of the spectrometer, which is placed 
slightly less than one focal length away, yielding a magnification of about 
6 for the image of the laser scattering region on the spectrometer slits.  
The spectrometer is a Jarrel Ash, model 82-000,
single grating spectrometer with a focal length of 50 cm. It is equipped with a
1800 g/mm holographic grating blazed at 500 nm.  We direct the primary laser path
vertically through the cell so that it would be parallel to the entrance slit, 
for the optimization of the scattered light collection into the
spectrometer. A 1/2$''$ Hamamatsu R647P PMT is attached to the spectrometer 
exit slit for light detection, and 
is operated in single photon counting mode.

\begin{figure}
\begin{center}
\includegraphics[width=2.9 in]{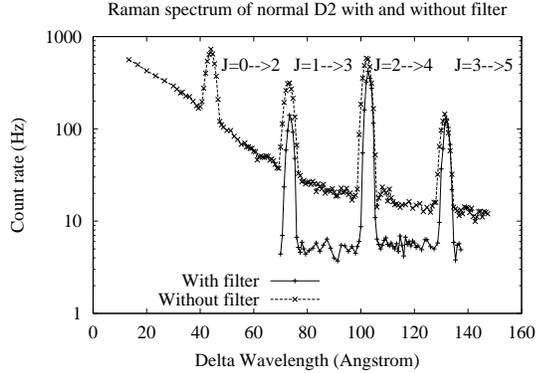}
\end{center}
\caption{Rotational Raman spectrum with and without a Raman filter.}
\label{figu_Ramanfilter}
\end{figure}

A 488 nm narrow band filter (bandwidth = 30 {\AA}), Omega Optical 
model XR3000-495AELP, placed in front of the spectrometer entrance,
introduces a crucial improvement on the performance of this system. 
This filter compensates for the poor stray light rejection of the 
single grating spectrometer,
and greatly suppresses the background counts in the spectrum, 
as shown in Fig. \ref{figu_Ramanfilter}.  
The price we paid for using such a filter is it
eliminates the first 
rotational peak, and distorts the second peak. Nevertheless, we
used the $J=2\rightarrow 4$ and $J=3\rightarrow 5$ peak to extract 
the ortho- to para-D$_2$ ratio with this spectrometer system.
For the bright $J=2\rightarrow 4$ rotational Raman
line in D$_2$, we observed roughly 1000 Hz at the peak of the line.

\begin{figure}[t]
\begin{center}
\includegraphics[width=2.9 in]{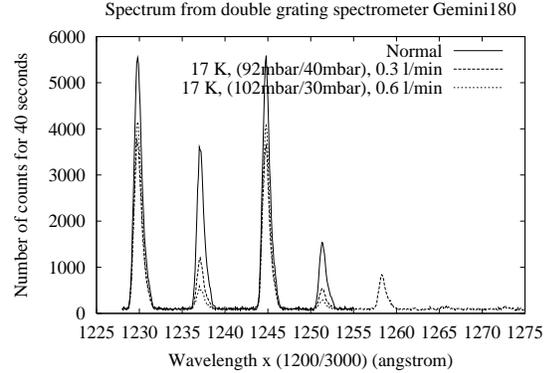}
\end{center}
\caption{Raman spectrum taken with the double grating spectrometer}
\label{figu_double}
\end{figure}

Some of our data was obtained using a double-grating system, ISA Gemini 180, with
a focal length of 180mm.  The optical
throughput for this system was a factor of two to three less than the single
grating system, but required no Raman filter (and therefore all four of the
lowest lying Raman rotational lines can be used).  
A sample spectrum from the double grating system is 
displayed in Fig. \ref{figu_double}.
Ultimately, because the
background was limited by the dark count rate in the PMT, the single grating
system was superior and was adopted for this application.

\begin{figure}[t]
\begin{center}
\includegraphics[width=2.9 in]{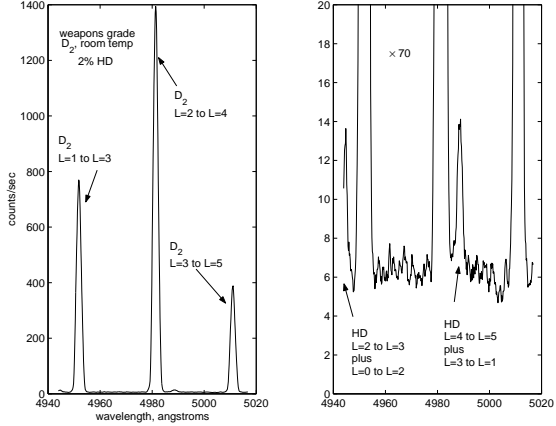}
\end{center}
\caption{D$_2$ Raman spectrum from the improved single grating spectrometer system.}
\label{figu_steve}
\end{figure}

Our final improvement was made, after these data were obtained, by utilizing
a thinner window between the camera lens and the focus of the laser, to reduce
distortions due to refraction of the scattered light, and a convex mirror to
direct the exiting laser beam back through the cell for a second pass.  These
improvements resulted in roughly a 30\% increase in the light collection
efficiency and a signal to background ratio of up to about 250:1 for gas
samples at atmospheric pressure and room temperature (see Fig. \ref{figu_steve}). 

\subsubsection{Results and Analysis of Raman Spectroscopic Measurements}

\begin{figure}
\begin{center}
\includegraphics[width=2.9 in]{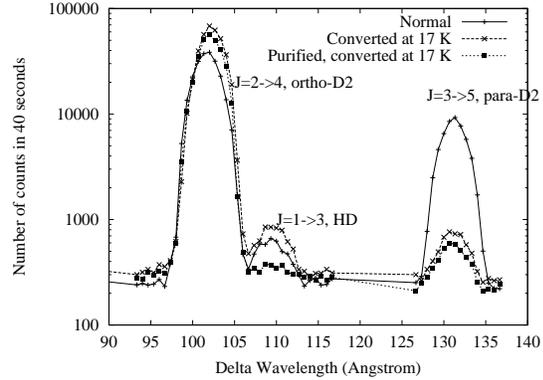}
\end{center}
\caption{The Raman spectrum of converted and unconverted D$_2$. Slit widths are set to
400 $\mu$m. The `+' curve is the reference normal D$_2$. The `$\times$' curve is a 
converted D$_2$ with (1.54\% $\pm$ 0.21\%) of para-D$_2$, and (1.27\% $\pm$ 0.12\%) of HD.
The square-dotted curve is purified converted D$_2$ with (1.42\% $\pm$ 0.20\%) of para-D$_2$, 
and a reduced HD contamination of (0.26\% $\pm$ 0.06\%).}
\label{figu_Ramanconvert}
\end{figure}

Our strategy to obtain the para-D$_2$ concentration using the rotational Raman
spectroscopy is based on a comparison between spectra of the sample with an unknown
ortho- to para-D$_2$ ratio and a reference with a known ratio. The measurements 
were carried out with samples at room temperature (T $\sim$ 300K). Typical spectra
obtained from our setup are shown in Fig. \ref{figu_Ramanconvert}.  According
to  Eq.(\ref{eq_ramansum}), the amplitude ratio of the 
$J=3 \rightarrow 5$ to the $J=2 \rightarrow 4$ peak among the Stokes
lines scales with the ratio of para-D$_2$ fraction, $f_p$, to the ortho-D$_2$
fraction, $f_o$, as  
\begin{equation}
\frac{\textrm{Amp}_{J=3 \rightarrow 5}}{\textrm{Amp}_{J=2 \rightarrow 4}} = \bigg(\frac{f_p}{f_o}\bigg) 
\frac{Q_p^{-1}g_3 7 e^{-E_3/kT} \sigma_{J=3\rightarrow 5}}
{Q_o^{-1}g_2 5 e^{-E_2 /kT} \sigma_{J=2 \rightarrow 4}}.
\end{equation}
\noindent Here the partition function $Q_p$ is a sum of spin multiplicities
weighted by the Boltzmann factor over all the odd rotational states of para-D$_2$, i.e., 
\begin{equation} 
Q_p=\sum_{odd \mbox{ }J} g_J (2J+1) exp(-E_J/kT),
\end{equation}
\noindent and for $Q_o$ the sum is over even $J$'s.
When compared with a spectrum of the reference sample measured at the same temperature
(room temperature), the statistical weighting factors and the cross sections are canceled out, leaving only the para to ortho ratio, $f_p/f_o$, of the unknown sample.
This procedure also cancels out the light collection efficiencies, making extremely
high precision measurements (limited by laser stability and knowledge of the reference
sample).  ``Normal'' D$_2$ with room temperature equilibrium population
of rotational states ($f_p/f_o$ = 1/2) is conveniently used as the reference.

The spectra in Fig. \ref{figu_Ramanconvert} illustrate the effect of para- to ortho-D$_2$
conversion.  The shrinkage of the $J=3 \rightarrow 5$ peak height of the 
converted D$_2$ sample indicates a reduction of the
para-D$_2$ concentration from 33.3\% (`+' curve, with D$_2$ collected
before the converter) to 1.54\% (``$\times$'' curve, with D$_2$ collected after the
17 K converter).  These spectra also show a noticeable HD peak ($J=1\rightarrow 3$) next
to the $J=2 \rightarrow 4$ ortho-D$_2$ peak.  

Because the UCN lifetime in solid D$_2$ is
also sensitive to the presence of hydrogen, which strongly absorbs UCN via nuclear capture
on the H nucleus, our knowledge of the HD concentration is also crucial in our study of the
superthermal S-D$_2$ source.  Rotational Raman spectroscopy provides 
a clean and powerful hydrogen mass spectrometer capability.  
To clearly identify HD rotational Raman peaks, a minimum peak
resolution of 5{\AA} is required.  After calibrating with a known concentration of HD, it
is possible to produce very sensitive tests of the presence of HD. The spectra shown in
Fig. \ref{figu_Ramanconvert} demonstrate the power in resolving a HD concentration to smaller
than 1\%. (In contrast to the spectra taken with D$_2$ sample of 2\%
HD, the square-dotted spectrum of a high purity, converted D$_2$ shows a 
reduced degree of HD contamination of 0.26\%.) 

Limited by the power of the laser, our Raman spectroscopy technique required
the pressure of the sample gas to be higher than roughly half an atmosphere to perform
para-D$_{2}$ peak identifications in 40 minutes with our first implementations of the
spectrometer.  Subsequent improvement of the light collection and 
replacement of our laser 
dropped this time to roughly 10 minutes.  The resolution of the para-D$_2$ peak
is limited by the background count rate (ideally the dark count rate of the photon
counting device). The PMT has a nominal dark count rate of 5 Hz at room temperature,
and with a para-D$_2$ peak count rate of 10 Hz it gives a relative uncertainty of
\begin{equation}
0.67 \frac{\sqrt{N_{pk}+N_{bg}}}{N_{pk}-N_{bg}} = 0.67 \frac{\sqrt{(10+5)t}}{(10-5)t}
\end{equation}
\noindent of the para-D$_2$ peak amplitude. A 30 second signal collection 
gives a measurement of the peak with a 10\% relative
uncertainty. For the case of highly converted samples,
a photon count rate of 1000 Hz under the ortho peak would give an absolute uncertainty of 0.15\% in
the knowledge of ortho-D$_2$ purity.   Comparable count rates have been achieved
with the setup described previously.

For spectra taken with the single grating spectrometer, 
we extracted the number of counts in each
Raman peak by fitting the peak to a Gaussian function plus a constant 
background. Using this functional form for the spectral peaks permitted us
to extract, in a consistent fashion, the area of HD peaks which sat on the
wings of more intense D$_2$ lines.
For the small para-D$_2$ peaks, these fits exhibited $\chi^2$ consistent with 
Poisson statistics, however, for the large ortho-D$_2$ peaks, the uncertainties
extracted from the fit were expanded to account for small discrepancies between the
peak shape and our Gaussian model.

To check our fitting procedure, we compared our results to those obtained by fitting
to a linear background region bracketing a Raman peak, and then simply subtracting
this background from the peak area. The linear fits to the background were consistent
with Poisson statistics, and the extracted ortho-to-para ratios were consistent with
our nominal fitting procedure. We took this as an indication that our quoted ratio
uncertainties were reasonable.

Note that the line extraction
procedure was slightly different for the double-grating spectrometer, due to a small distortion
in the line shape (later corrected by adding a baffle to the first stage of the double-grating
spectrometer).  We adopted a conservative approach of expanding the fitting window around the
line and accepting a somewhat poorer peak-to-background ratio to extract these ratios.
The uncertainties listed here are of 95\% confidence level. 

The extraction of the
para-D$_2$ fraction is based on the assumption of full conversion of the room temperature
D$_2$ reference sample.  The fact that we use purified D$_2$ flowing through a hot
(400 $^{\circ}$C) palladium purifier, which is a highly active ortho/para-D$_2$ catalyst, gives
us some confidence in this assumption.  Note also that the ratios of the populations reproduce
the expected thermal populations when the appropriate absolute Raman cross-sections for the
different, measured Raman lines is utilized.  

\begin{figure}
\begin{center}
\includegraphics[width=2.9 in]{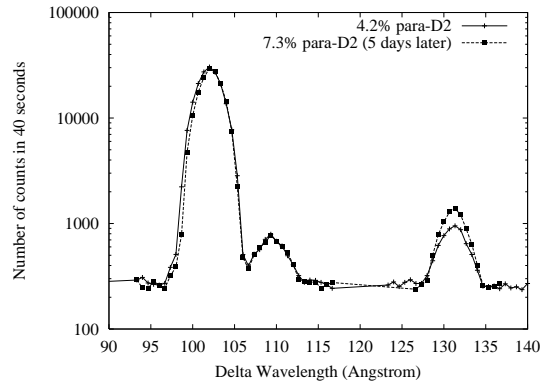}
\end{center}
\caption{Relaxation in the Raman cell: The para-D$_2$ concentration grows from
4.2\%(`+' set), converted at 23.6K, to 7.3\%(dotted set) over a 5 day period.}
\label{figu_Ramanrelax}
\end{figure}

The precision of the
para-D$_2$ fraction is currently statistically limited by the signal to background ratio
of the small para-D$_2$ peak of the converted sample.  It can be improved with a higher
D$_2$ gas pressure (perhaps using a compressor), a stronger laser source, or reducing
the present dark count rates in our spectrometer's PMT.
Ultimately, systematic corrections due to relaxation, laser power fluctuations,
temperature fluctuation, 
{\it etc...}, limit the precision of this technique.

The effect of relaxation in the sample cell was also investigated.
As mentioned, this optical technique requires gas samples with 
pressures of several hundred Torr.  During the conversion/fill stage,
the line pressure of the converted D$_2$ is kept under 
50 Torr; without introducing another means of volume compression, our only technique for collecting gas samples of several hundred Torr is 
from the boil-off of the condensed D$_2$ in the end of each run cycle when
the cryostat is warmed up and the D$_2$ is removed from the cryostat in the form
of vapor. 
After these samples are collected, there was usually several hours of delay from the
time of sample collection to the time of measurement.  This delay raises
the possibility of significant relaxation of ortho- into para-D$_2$ in
our room temperature optical sample cell, which has S.S. surfaces.
Fig. \ref{figu_Ramanrelax} shows the observable relaxation over the course of 5 days.
This corresponds to an increase of 0.3\% of para-D$_2$ concentration overnight.
Relaxation of ortho-D$_2$ introduces a correction within the measurement uncertainty
for samples measured a few hours after the collection. When higher precision is
required, a non-magnetic optical cell might be required.

In conclusion, we have demonstrated the technique of Raman spectroscopy to perform
high precision measurements of the ortho-D$_2$ purity to 0.15\%, using a
low power visible laser and a simple, single-grating spectrometer with
an optimized light throughput. This provides us with the capability of cleanly
determining the absolute para-D$_2$ concentration, as well as the level of HD
contamination in D$_2$ sample obtained from our source. The fact that
this technique requires almost no calibration makes it superior to a 
conventional scheme which we discuss below.

\subsection{Thermal Conductivity Measurements -- Hot Wire Method}

The hot wire method is an application of the traditional scheme to 
measure the 
thermal conductivity of various gases\cite{Tsederberg}. It utilizes a 
thin metal wire as the heat source as well as the temperature 
sensor.
The supplied energy heats up the wire, and the produced 
heat is conducted toward the cold walls
through the surrounding medium (in our case, D$_2$ gas). 
The rate of change as well as the ultimate temperature of the wire 
depends on the macroscopic thermal properties of the gas medium, and 
thus a measurement of the wire resistance (by measuring 
the voltage across the wire) 
is expected to be sensitive to the concentration of para-D$_2$. 

We built a system, similar to the design in Ref.\cite{Stewart55,Grilly53}, of two cylindrical cells
with a 1$''$ I.D. and a 5.5$''$ height out of one rectangular cubic block of OFHC copper. 
Constantan conductor feedthroughs in ceramic insulators were used for the electrical
leads.  Tungsten wires of 0.5 mm O.D. were used as the hot wires and were soldered to
the Constantan leads. The wires run along the cylindrical axis of the cell and
are slightly stretched by springs attached to the 
bottom of the cell.  The whole assembly
is immersed in a liquid N$_2$ bath. The outer cylindrical surfaces are
maintained at a constant temperature of 77K. 
The reference cell was filled with 30 mbar of $^4$He,
and the other cell with 30 mbar of D$_2$.

\begin{figure}
\begin{center}
\includegraphics[width=2.9 in]{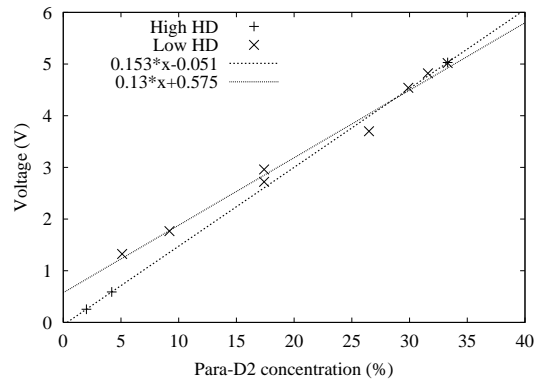}
\end{center}
\caption{The performance of the hot wire cell.
The `+' data set is taken with a unpurified D$_2$ sample (with a high HD concentration of
2.05\% $\pm$ 0.16\%) The knowledge of the para-D$_2$ concentration 
obtained from the Raman spectroscopy. The ``$\times$'' set of data is taken with gas
mixtures with known volume ratio of the converted D$_2$ at 17 K and the normal D$_2$.
HD contamination is approximately 0.3\%.
}
\label{figu_TCdata}
\end{figure}

Our voltage data showed the expected behavior $vs.$ para-D$_2$ fraction,
where the scatter around the linear dependence sets the precision in the para-D$_2$
fraction to be 2.6\% (we view this as a preliminary figure, with substantial
improvements possible with a more sophisticated design for the conductivity
measurement apparatus\cite{Devoret80,Haarman71,Wakeham81}). A systematic
effect was found that 
HD contamination changes the slope of the voltage data set.
The effect of the HD presence on the voltage measurements is
consistent with HD having slightly larger specific heat
and viscosity than ortho-D$_2$ at the temperature range
of operation.  Ultimately the HD concentration
should be quite low in an UCN source operated in ``production mode,'' to limit UCN
losses via nuclear absorption on H contamination, however, without an independent
method of evaluating this contamination $in\ situ$, it would appear that the
hot wire cell technique would produce misleading results.

\section{Discussions of the Converter Performance}

With these tools to monitor the para-D$_2$ concentration, we recorded the performance
of the cryogenic converter containing a hydrous ferric oxide catalyst, over nine months
of intermittent source operation.  The catalyst was not regenerated until after the
last of these source studies, a series of higher UCN flux runs, in which the UCN
lifetime had shortened significantly.

\begin{figure}
\begin{center}
\includegraphics[width=2.9in]{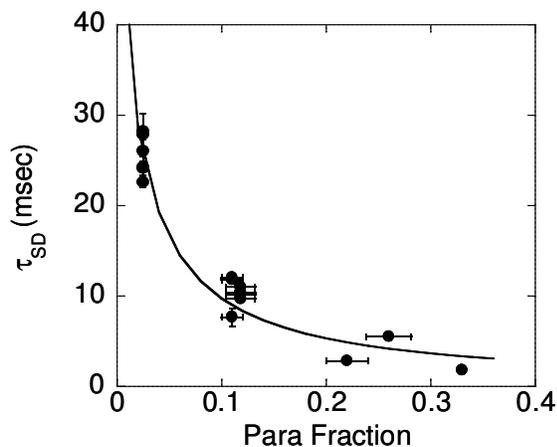}
\end{center}
\caption{Neutron lifetime as a function of the para-D$_2$ fraction in the solid
deuterium.\cite{Morris01}}
\label{figu_neutron}
\end{figure}

The best level of conversion gave a 1.4\% $\pm$ 0.2\% para-D$_2$ concentration.
These results were achieved with a converter temperature of 17K, an outlet pressure
at the converter of around 35 mbar, and a flow rate of around 0.5 STP liters/min.
The conversion efficiency of the catalyst reduced over time,
with the 17K converter producing
converted D$_2$ gas with a para-D$_2$ fraction growing from 1.5\% to 11\% over
the period in which our UCN experiments were conducted. These
measurements were performed to determine the UCN lifetime in solid deuterium. 
In Fig. \ref{figu_neutron}, we present the measured UCN lifetimes as a function of
the para-D$_2$ fraction.\cite{Morris01} The extracted value for the lifetime
in a pure para-D$_2$ crystal is in
excellent agreement with the theoretical expectations and provides
substantiating evidence that our technique is effective.

Another point of contact between the function of the converter and our two para fraction
measurement techniques can be made through tests conducted at the end of our measurement period.
Although the efficiency of the converter had deteriorated, we investigated the converter
performance by adjusting (via. the attached heater) the temperature of the converter cell
and then measuring the resultant para- to ortho-D$_2$ ratios immediately.  Once again, we were
able to produce para- to ortho-D$_2$ ratios at the 1.5\% level, although for these measurements the converter was
held at 16 K and the flow rate was reduced.

The problem of catalyst aging has been well investigated in literatures.\cite{Weiser35}  Hydrous gels lose
water and coalesce into granules over time. This change of structure from amorphous gel to
micro-crystals is evident from X-ray scattering experiments\cite{Weiser35}.  The volume
shrinks by over a factor of 500 due to this dehydration, and results in
the reduction of the conversion efficiency.  It is thus advisable to work
with fresh gels, and not to attempt repeated regenerations.  One way to prevent dehydration
might involve refrigeration of the catalyst right after formation.  
Regeneration of the catalyst by heating it to 130$^{\circ}$C under atmosphere was
reported to restore the catalytic performance\cite{Weitzel58}, however, for the intermittent,
brief use of the converter we envision for our source, one might as well replenish the converter
with fresh catalyst when necessary. One potential concern
regards the possible formation of HD while  D$_2$ gas passes through the hydrogen-rich
catalyst via isotope exchange.  This, however, was not experimentally observed 
probably because of the low operational temperature of the catalyst cell.

At present, the dependence of converter efficiency on its various operational parameters
(temperature, pressure, flow rate, etc...) remain to be thoroughly investigated.  At least
two effects are worthy of note.  We observed that the vapor pressure at the outlet of
the converter was roughly a factor of ten smaller than one might expect for D$_2$ in equilibrium
with its saturated vapor pressure at the converter cell temperature (up to 30K).
One would intuitively attribute this phenomenon to the large binding energy of
the catalyst surface.
We were able to flow
adequate quantities of D$_2$ through the cell by maintaining a pressure gradient of 50 to 80
mbar across the cell, but when the cell was operated at temperatures below 21K we
occasionally
experienced a gradual increase in the flow impedance, which may have been the result of forming
solid D$_2$ ``plugs'' due to the shift of the gas-solid phase transition curve. 
This plugging phenomenon may also be attributable to water contamination of
our D$_2$ gas. In fact
we also observed that pre-purifying the deuterium gas through the palladium 
filter was essential in preventing the ice plugs from forming inside the 
para-to-ortho converter.
 
When scanning the temperature of our conversion
cell across the triple point, we measured higher para fractions than were expected
near the phase transition temperature.  This phenomenon was observed in two separate
scans through the triple point, with both the Raman spectroscopy and thermal conductivity
cell measurements.  These data appear to be consistent with the
latent heat of a phase transition in the D$_2$ providing the necessary energy to repopulate
the J$=1$ rotational state, but given the limited quantity of data we have on 
this phenomenon at present, we feel more work is required before a compelling conclusion can be drawn.  In particular, these traces were obtained at the end of the 9 month
operation period in which the converter material was not regenerated, when the performance
of our converter was significantly impaired and may not represent typical operating conditions.

In conclusion, we have produced highly converted D$_2$ with a residual para-D$_2$ concentration of 1.5\%. 
This is done with a 100 c.c. converter filled with hydrous ferric oxide operated
at 17 K, with a flow rate of roughly 0.5 liter/min.  We developed an apparatus to monitor
para-D$_2$ and HD contamination to a high precision at levels of 2 parts in 10$^3$ and 1
part in 10$^3$, respectively.  Our study of the functionality of the converter leads us to
believe that regular replacement with fresh catalyst or thorough regeneration will be
required to ensure adequate para-to-ortho-D$_2$ conversion and that a catalyst which does not
largely reduce the D$_2$ vapor pressure may help avoid complications associated with condensation.

\end{document}